# Superconductivity in Ag implanted Au thin film

Manas Kumar Dalai[1,2,*], Braj Bhusan Singh[3], Salila Kumar Sethy[1,2], Satya Prakash Sahoo[4,5], Subhankar Bedanta[3,**]

[1]CSIR – Institute of Minerals and Materials Technology, Bhubaneswar – 751013, Odisha, India

[2]Academy of Scientific and Innovative Research (AcSIR), Ghaziabad – 201002, India

[3]Laboratory for Nanomagnetism and Magnetic Materials (LNMM), School of Physical Sciences, National Institute of Science Education and Research (NISER), HBNI, Jatni, 752050, India

[4]Institute of Physics, Bhubaneswar – 751005, India

[5]Homi Bhaba National Institute (HBNI), Anushaktinagar, Mumbai 400 085, India

**Abstract:**

Au and Ag are known to be metallic and none of them have shown independently superconductivity. Here we show superconductivity in Ag implanted Au thin films. Ag implanted Au films of different thicknesses have been studied using four probe resistance versus temperature measurements. We have also measured the pristine samples (i.e. without any implantation) to compare the transport properties. The superconducting transition occurs around ~2 K ($T_C$) for Ag implanted 20 nm Au film (the lowest resistance observed below $T_C$ is ~$10^{-6}$ Ohm). Change of $T_c$ towards lower temperatures has been observed under the effect of increasing current and magnetic field which endorses the superconducting state. Further the $V{\sim}I$ measurements were carried out to verify the superconducting nature. On the other hand, the pristine samples clearly did not exhibit superconductivity down to 1.6 K. Our results open an avenue for further exploring this type of novel superconductivity for future applications.

**Introduction:**

Today the word "Superconductivity" gives a sparkling feeling to the materials scientists even after 100 years of its discovery by Heike Kamerlingh Onnes in Mercury [1]. Since its discovery, scientists are in race to find such novel phenomena at higher temperatures with a great aim towards the room temperature superconductors. The discoveries of superconductivity at higher temperatures in Lanthanum- and later on Yitrium- based cuprates have enlighten the community with large hope to achieve the superconductivity at room temperature [2,3].

In this context the search for high $T_C$ (above liquid nitrogen temperature) superconductors [Ref. 3 - 9] have drawn constant interest in last few decades. Further, focus also has been put to observe superconductivity in various new materials [10 -16].

The origin to the superconductivity state for some of the materials are well explained with the mechanism of conventional BCS theory [17,18] (conventional superconductor) and for some materials BCS theory does not explain (unconventional superconductor) [19,20,21,22]. There are a few materials for which the understanding of the exact origin of superconductivity is in progress [23,24,25]. From application point of view, superconductors have already been used in various technologies such as Cryotron, superconducting quantum interference device (SQUID), magnetic resonance imaging (MRI), nuclear magnetic resonance (NMR), single photon detector, etc. [26-30]. In particular, the recent research shown by Thapa and Pandey on superconductivity at ambient conditions of temperature and pressure in Ag nanoparticles in Au matrix has triggered researchers to dream for room temperature superconductors [31]. Reproducibility of this result by others is yet to come. Rather, Biswas et. al has reported the absence of superconductivity in Ag-Au modulated nanostructured thin films [32]. On the other hand G. Baskaran has recently made theoretical calculation and suggests the existence of superconductivity in Ag-Au system [33]. It is also noted that superconductivity has been reported in various Au alloys at low temperatures [34, 35], however, not in Ag-Au system except the report by Thapa and Pandey [31]. In this context we have investigated transport property of Ag implanted Au films which clearly shows the existence of superconductivity.

**Methods**

Two different thickness of Au films (20 and 5 nm) were deposited using e-beam evaporator in a high vacuum chamber manufactured by Mantis Deposition Ltd. (Model: QPrep 500) on Si (100) substrate having a native $SiO_2$. A thin layer of Ta was used as a buffer layer.

Prior to deposition, the Si substrate was cleaned in acetone and isopropanol in ultrasonic for five minutes. For all the Au films, a 3nm thick Ta was deposited by dc sputtering on Si substrate as a buffer layer. The base pressure in the chamber was kept ~ 7 x $10^{-8}$ mbar. For uniform film thickness, the substrate was rotated at 20 rpm while depositing Ta and Au. All the thin films were deposited at room temperature.

Ag-ion implantation in Au films were carried out using a 50 KeV low energy ion implanter at Institute of Physics, Bhubaneswar. The pristine samples (i.e. without any implantation) are named as S1 and S2 for Au thickness of 20 and 5 nm, respectively. Further the Ag implanted samples are named as S1-Ag and S2-Ag for Au thickness of 20 and 5 nm, respectively. The energy of ion (Ag) beam was kept different such as 30 KeV and 10 KeV for samples S1-Ag and S2-Ag, respectively. Resistance vs temperature (*R-T*) measurements in the temperature range of 1.6 to 300 K were carried out using four probe (van der Pauw) technique in a low-temperature cryostat manufactured by Oxford Instruments (SpectromagPT) at variable magnetic fields (0 to 1 Tesla) and currents (0.05 to 4.00 mA). A current source (Keithley 6220) and a nanovoltmeter (Keithley 2182A) were used to apply current and measure voltage, respectively. Further we have measured the *V-I* characteristics at various temperatures down to 1.6 K (lowest temperature achievable in the SpectromagPT cryostat, which was used for this experiment).

**Results and Discussion**

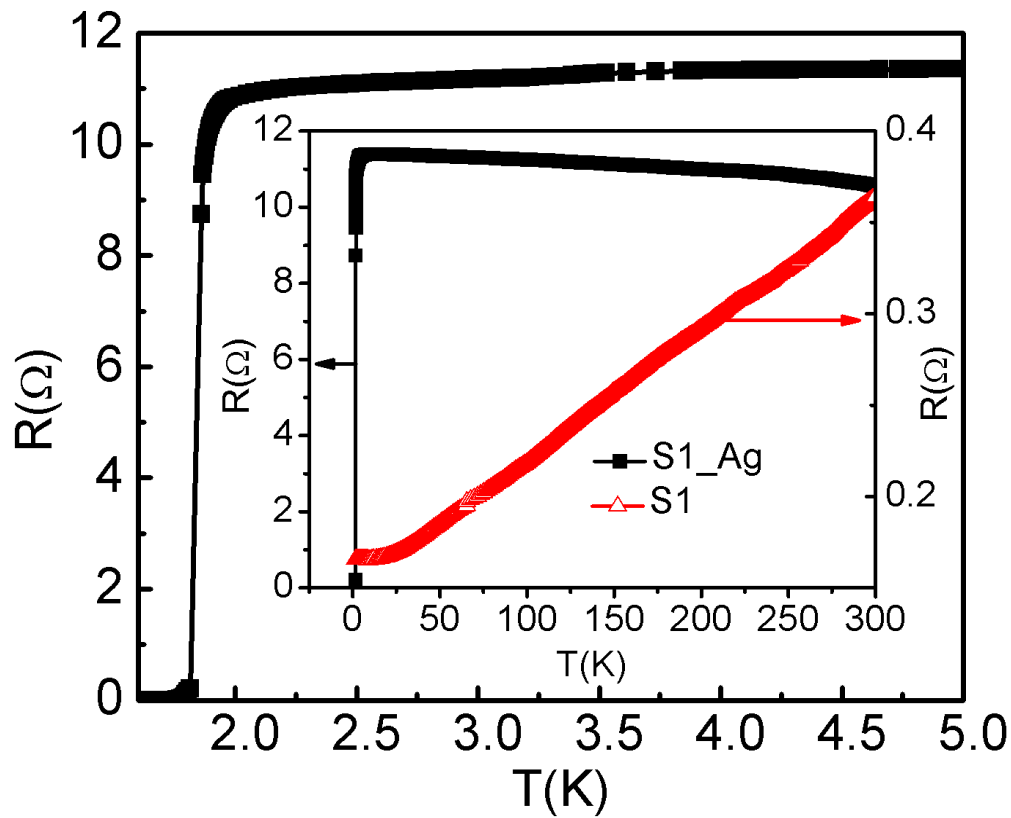

Figure 1: Resistance versus temperature graph for the Ag-implanted 20 nm Au film (sample S1-Ag). The insert shows the *R-T* behaviour in the extended temperature range of 1.6 to 300 K for both the samples S1-Ag and its corresponding pristine sample S1 . All the data have been measured at $I = 0.1$ mA.

Ag implanted samples are named as S1-Ag and S2-Ag for Au thickness of 20 nm, and 5 nm, respectively. In the following we will discuss mostly the results obtained from sample S1-Ag. Fig. 1 shows the resistance (*R*) versus temperature (*T*) plot for sample S1-Ag measured at $I = 0.1$ mA. It is clearly observed that the resistance value starts decreasing below ~ 4 K and a sudden fall is seen at ~ 1.86 K. The inset of Fig. 1 shows *R* vs *T* behaviour in the extended temperature range (1.6 to 300 K). The lowest resistance value of this sample drops to ~$10^{-6}$ Ohm below ~ 1.86 K. It is a clear signature of superconductivity in sample S1-Ag at around 1.86 K, hence, thereafter called as the superconducting transition temperature $T_C$. The $T_C$ value was extracted by differentiating the *R-T* curve. We have also performed the *R-T* measurements (red coloured triangles shown in inset of Fig. 1) of pristine sample S1, where no such superconductivity transition was observed.

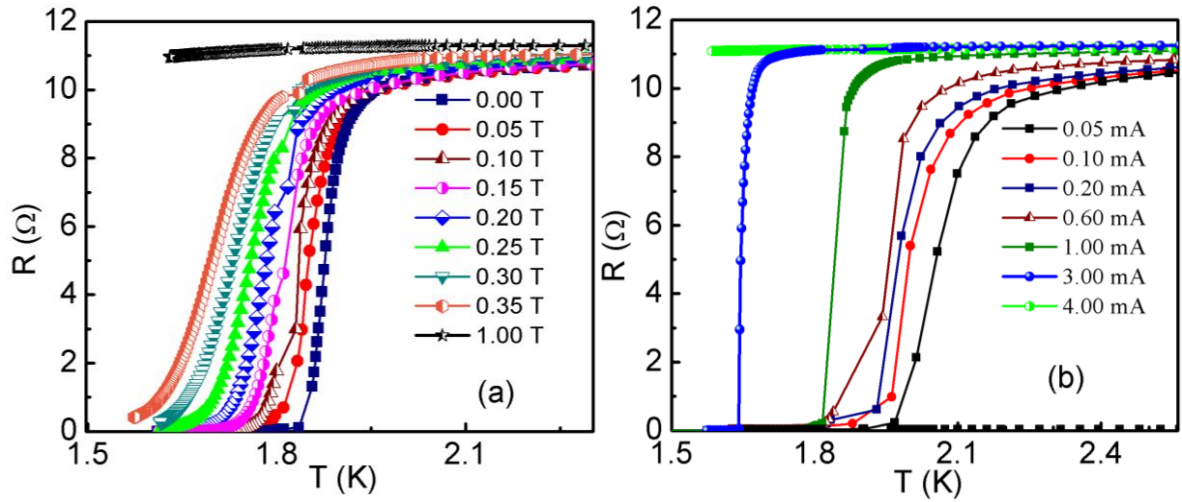

Figure 2: (a) Resistance versus temperature for the implanted sample S1-Ag measured at different magnetic fields and at a constant current of $I = 0.1$ mA. (b) Resistance versus temperature for the same sample S1-Ag measured at different current.

To further verify this novel effect, we have carried out the *R-T* measurements under various magnetic fields (Fig. 2(a)) and at a constant current of $I = 0.1$ mA for sample S1-Ag. It is seen that at zero magnetic field the *Tc* is found to be ~1.87 K. The *Tc* values extracted from Fig. 2 for magnetic fields of 0, 0.05, 0.1, 0.15, 0.2 and 0.25 Tesla are 1.87, 1.85, 1.84, 1.83, 1.78 and 1.76 K, respectively. Hence *Tc* shifts towards lower temperatures for increasing fields which is a typical behaviour of superconductive materials [36-38]. In Fig. 2(b) we have presented the current dependence nature of *R-T* curves for various currents (0.05, 0.1, 0.2, 0.6, 1.0, 3.0, and 4.0 mA). The *Tc* was found to be 2.06 K for $I = 0.05$ mA. There is a clear change in the $T_c$ shifting to lower temperatures (like other superconducting systems [39]) when we increase the current from 0.05 mA to 3 mA and with further increase of current to 4 mA, the transition is not observed within our temperature limit of the cryostat.

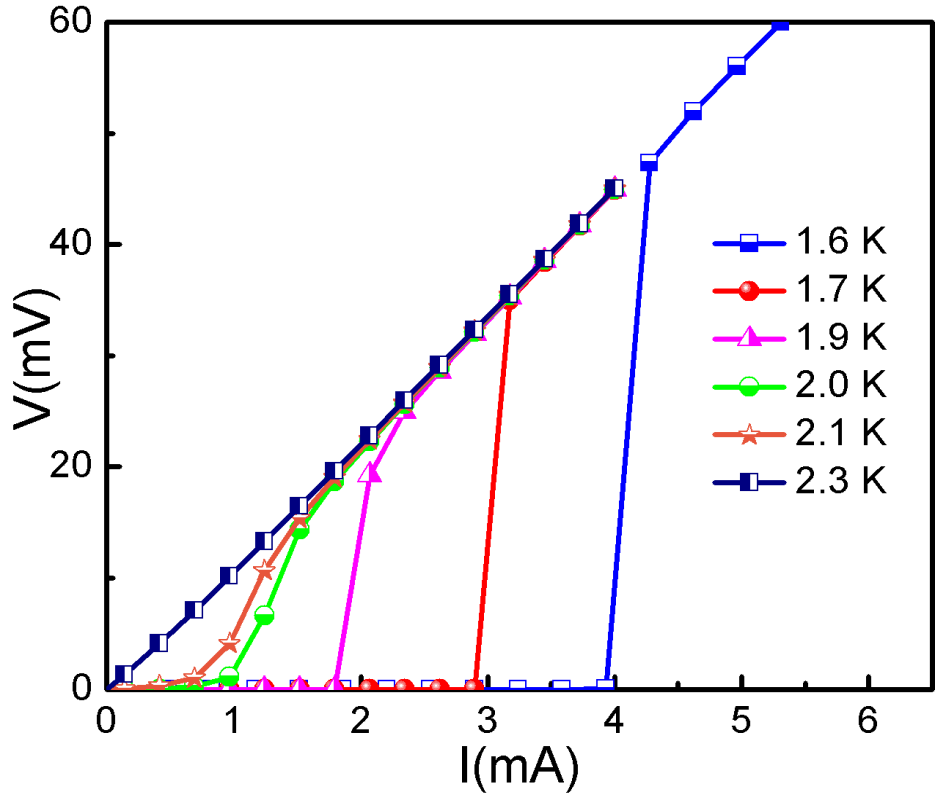

Figure 3: Voltage (*V*) vs current (*I*) measurements at different temperatures around *Tc* for the implanted sample S1-Ag.

Further we have carried out the voltage (*V*) versus current (*I*) measurements of S1-Ag sample at various temperatures around the *Tc* as presented in Fig. 3. At 2.3 K, the voltage varies linearly (Ohmic behaviour) with current and the line passes through the origin. However, by further decreasing the temperature the Ohmic (linear) behaviour transforms to superconducting behaviour (plateaus in *V-I* plot). It should be noted that the width of the plateaus increases with decrease in temperature below the *$T_C$*. This indicates that below $T_C$ the critical current goes on increasing with decrease of temperature as reported earlier in other superconductors. [37,38]

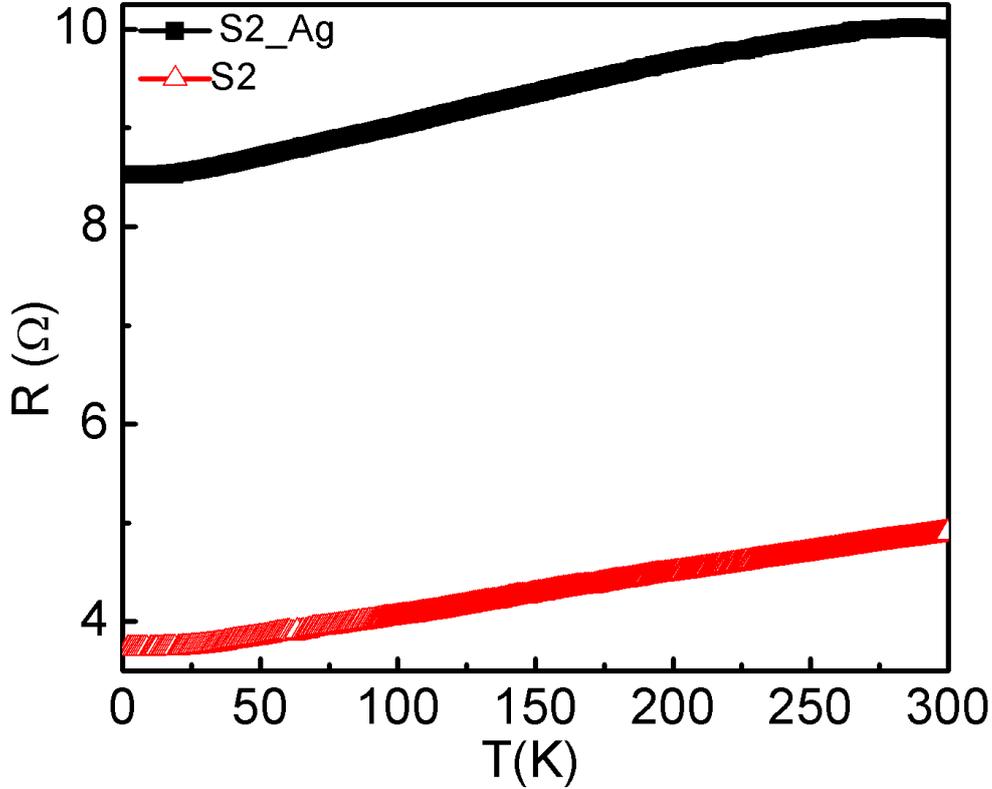

Figure 4: Resistance (*R*) versus temperature (*T*) of pristine (S2) and Ag-implanted (S2-Ag) 5 nm Au film represented by triangle (red) and square (black) symbols, respectively.

To further analysis of transport behaviour in different thickness of Au films, we have also undertaken the *R-T* measurements of Ag implanted 5 nm Au film (S2-Ag) as well as its corresponding pristine sample S2 are presented in Fig. 4. In sample S2-Ag, we could not observe the transition until our lowest system temperature limit, i.e., 1.6 K. The pristine sample (S2) shows usual metallic behaviour.

Room temperature superconductivity in Ag-Au system is recently reported by Thapa and Pandey [31]. They studied a system of Ag nanoparticles embedded in Au matrix. Several research groups took initiative to observe such novel effect in Ag-Au system. For example, Biswas et al.[32] has attempted to observe superconductivity in Ag-Au thin films prepared by pulsed laser deposition. However, their studied system did not exhibit any superconductivity in the temperature range of 5 to 300 K. On the other hand, theory groups have also attempted to verify the superconducting behaviour in Ag-Au system. Singh et al.[40] has made first principle calculations on 3D bulk crystals and 2D slabs of Ag-Au binary alloys where the $T_C$ was calculated to be as low as 1mK. Further the recent theoretical work by Baskaran endorses the superconductivity in Ag-Au system [33]. In his work, it is suggested that electron transfer may occur from Ag to Au matrix at the Ag-Au interface and quasi 2D structural reconstruction can happen leading to confined superconductivity.

**Conclusions:**

In conclusion, we have reported the existence of superconductivity in Ag-implanted 20 nm Au thin film. The *Tc* was found to be ~ 2 K below which the resistance was measured to be ~ $10^{-6}$ Ohms. Further the magnetic field and current dependent resistance versus temperature confirms the superconducting behaviour. Our results pave a path to explore the new paradigm of superconductivity in Ag-Au system involving ion-implantation. Future work should be focused to study the effect of various dose and/or energy of the implanted Ag ions in different thicknesses of Au matrix. It may be possible to increase the $T_C$ in this system by varying the growth conditions. Both theory and experimental work on this topic will bring new insights into the origin of this novel superconductivity.

**Acknowledgement:** BBS acknowledges DST, Govt. of India, for the INSPIRE faculty fellowship. SKS acknowledges the Council of Scientific and Industrial Research (CSIR), India, for fellowship. We thank Mr. Anup Kumar Behera and Mrs. Ramarani Dash of Institute of Physics, Bhubaneswar, for their help during ion-implantation experiment. Our sincere thanks to Prof. G. Baskaran of Institute of mathematical sciences, Chennai, for valuable discussions. We acknowledge department of atomic energy (DAE), Govt. of India, for establishing various experimental facilities used in this work.

Correspondence: dalaimk@immt.res.in (Manas Kumar Dalai)

sbedanta@niser.ac.in (Subhankar Bedanta)

References:

1. H. K. Onnes, Commun. Phys. Lab. Univ. Leiden. 12, 120 (1911)
2. J. G. Bednorz and K. A. Müller, Z. Phys. B. 64 (1), 189 (1986)
3. M. K. Wu, J. R. Ashburn, C. J. Torng, P. H. Hor, R. L. Meng, L. Gao, Z. J. Huang, Y. Q. Wang, and C. W. Chu, Physical Review Letters. 58 (9), 908, (1987).
4. A. Schilling, M. Cantoni, J. D. Guo and H. R. Ott, Nature, 363, 56 (1993)
5. P.Dai, B. C.Chakoumakos, G.F.Sun, K.W.Wong, Y.Xin and D.F.Lu, Physica C, 243, 201 (1995)
6. A. P. Drozdov, M. I. Eremets, I. A. Troyan, V. Ksenofontov and S. I. Shylin, Nature, 525, 73 (2015)
7. Z. Z. Sheng and A. M. Hermann, Nature, 332, 55 (1988)
8. Donglu Shi, Mark S. Boley, J. G. Chen, Ming Xu, K. Vandervoort, Y. X. Liao and A. Zangvil, Appl. Phys. Lett. 55, 699 (1989)
9. Jian-Feng Ge, Zhi-Long Liu, Canhua Liu, Chun-Lei Gao, Dong Qian, Qi-Kun Xue, Ying Liu & Jin-Feng Jia, Nature Materials, 14, 285 (2015)
10. J. Shiogai, Y. Ito, T. Mitsuhashi, T. Nojima and A. Tsukazaki, Nature Physics, 12, 42 (2016)
11. M. Mitrano, A. Cantaluppi, D. Nicoletti, S. Kaiser, A. Perucchi, S. Lupi, P. Di Pietro, D. Pontiroli, M. Riccò, S. R. Clark, D. Jaksch and A. Cavalleri, Nature, 530, 461 (2016)

12. Alexey Y. Ganin, Yasuhiro Takabayashi, Yaroslav Z. Khimyak, Serena Margadonna, Anna Tamai, Matthew J. Rosseinsky and Kosmas Prassides, Nature Materials, 7, 367 (2008)
13. Joshua H. Tapp, Zhongjia Tang, Bing Lv, Kalyan Sasmal, Bernd Lorenz, Paul C. W. Chu, and Arnold M. Guloy, Phys. Rev. B 78, 060505 (2008)
14. Menghan Liao, Yunyi Zang, Zhaoyong Guan, Haiwei Li, Yan Gong, Kejing Zhu, Xiao-Peng Hu, Ding Zhang, Yong Xu, Ya-Yu Wang, Ke He, Xu-Cun Ma, Shou-Cheng Zhang & Qi-Kun Xue, Nature Physics, 14, 344 (2018)
15. Shingo Yonezawa, Kengo Tajiri, Suguru Nakata, Yuki Nagai, Zhiwei Wang, Kouji Segawa, Yoichi Ando and Yoshiteru Maeno, Nature Physics, 13, 123 (2017)
16. Yasuhide Tomioka, Naoki Shirakawa, Keisuke Shibuya and Isao H. Inoue, Nature Communications, 10, Article number: 738 (2019)
17. J. Bardeen, L. N. Cooper, and J. R. Schrieffer, Phys. Rev. 106, 162 (1957)
18. J. Bardeen, L. N. Cooper, and J. R. Schrieffer, Phys. Rev. 108, 1175 (1957)
19. H. R. Ott, H. Rudigier, Z. Fisk, and J. L. Smith, Phys. Rev. Lett. 50, 1595 (1983)
20. G. R. Stewart, Z. Fisk, J. O. Willis, and J. L. Smith, Phys. Rev. Lett. 52, 679 (1984)
21. T. T. M. Palstra, A. A. Menovsky, J. van den Berg, A. J. Dirkmaat, P. H. Kes, G. J. Nieuwenhuys, and J. A. Mydosh, Phys. Rev. Lett. 55, 2727 (1985)
22. A. Di Bernardo, O. Millo, M. Barbone, H. Alpern, Y. Kalcheim, U. Sassi, A. K. Ott, D. De Fazio, D. Yoon, M. Amado, A. C. Ferrari, J. Linder and J. W. A. Robinson, Nature Communications, 8, Article number: 14024 (2017)
23. Anthony J. Leggett, Nature Physics, 2, 134 (2006)
24. D. van der Marel, J. L. M. van Mechelen, and I. I. Mazin, Phys. Rev. B 84, 205111 – (2011)
25. Lev P. Gor'kov, PNAS April 26, 113 (17), 4646 (2016)
26. D. A. Buck, Proceedings of the IRE. 44 (4), 482 (1956)
27. R. C. Jaklevic; J. Lambe; A. H. Silver and J. E. Mercereau. Physical Review Letters. **12** (7): 159 (1964)
28. M. C. K. Wiltshire, J. B. Pendry, I. R. Young, D. J. Larkman, D. J. Gilderdale, and J. V. Hajnal, Science 291(5505), 849 (2001)
29. Ya. S. Greenberg, Reviews of Modern Physics, 70, 175 (1998)
30. G. N. Gol'tsman, O. Okunev, G. Chulkova, A. Lipatov, A. Semenov, K. Smirnov, B. Voronov, and A. Dzardanov, Applied Physics Letters 79, 705 (2001)
31. Dev Kumar Thapa, Anshu Pandey, arxiv 1807:08572 (2018).
32. Abhijit Biswas, Swati Parmar, Anupam Jana, Ram Janay Chaudhary and Satish Chandra Ogale, arXiv:1808:10699 (2018)
33. G. Baskaran, arXiv:1808:02005 (2018).
34. D.C. Hamilton, Ch.J. Raub, B.T. Matthias, E. Corenzwit and G.W. Hull, Jr.,J. Phys. Chem. Solids, 26, 665 (1965)
35. J.H. Wernick, A. Menth, T.H. Geballe, G. Hull and J.P. Maita, J. Phys. Chem. Solids, 30, 1949 (1969)
36. H. A. Leupold, F. Rothwarf, J. J. Winter, J. T. Breslin, R. L. Ross, T. R. AuCoin, and L. W. Dubeck, J. Appl. Phys. 45, 5399 (1974)
37. Ying Xing, Hui-Min Zhang, Hai-Long Fu, Haiwen Liu, Yi Sun,1 Jun-Ping Peng, Fa Wang, Xi Lin, Xu-Cun Ma, Qi-Kun Xue, Jian Wang, X. C. Xie, Science, 350, 542 (2015)

38. Yi Sun, Jian Wang, Weiwei Zhao, Mingliang Tian, Meenakshi Singh and Moses H. W. Chan, Scientific Reports, 3, Article number: 2307 (2013)
39. S. Mitra, A. P. Petrovi´, D. Salloum, P. Gougeon, M. Potel, Jian-Xin Zhu, C. Panagopoulos and Elbert E. M. Chia, Phys. Rev. B 98, 054507 (2018)
40. Surender Singh, Subhamoy Char, Dasari L. V. K. Prasad, arXiv:1812**:**09308 (2018)